\documentclass[12pt]{article}

\usepackage{graphicx}
\DeclareGraphicsExtensions{.ps,.eps,.pcx}
\usepackage{amsmath}
\usepackage{amssymb}

\usepackage{latexsym}
\usepackage{eucal}
\usepackage{color}
\usepackage{multirow}
\usepackage{enumitem}

\newtheorem{theorem}{Theorem}

\newtheorem{proposition}{Proposition}
\newtheorem{lemma}{Lemma}

\newcommand{\bbo}{\mbox{1}\hspace{-3pt}\mbox{I}}

\newcommand{\var}{\mbox{var\,}}

\newcommand{\cov}{\mbox{cov\,}}

\begin{document}

\thispagestyle{empty}

\centerline{\Large\bf A data-driven P-spline smoother}

\vspace{0.3cm}

\centerline{\Large\bf  and the P-Spline-GARCH models\footnote{ Financial support of the European Union's Horizon 2020 research and innovation program "FIN- TECH: A Financial supervision and Technology compliance training programme" under the grant agreement No 825215 (Topic: ICT-35-2018, Type of action: CSA), the European Cooperation in Science \& Technology COST Action grant CA19130 - Fintech and Artificial Intelligence in Finance - Towards a transparent financial industry, the Deutsche Forschungsgemeinschaft's IRTG 1792 grant, the Yushan Scholar Program of Taiwan, the Czech Science Foundation's grant no. 19-28231X / CAS: XDA 23020303, and the DFG project FE 1500/2-1 are greatly acknowledged. Both authors declare that they have no  conflict of interest.}}

\vspace{0.5cm}

\centerline{\large Yuanhua Feng$^a$\footnote{\vspace*{-0.2cm}Corresponding address: Yuanhua Feng, Paderborn University, Warburgerstr. 100, D-33098 Germany. \\ \hspace*{0.65cm}Email: yuanhua.feng@uni-paderborn.de; Tel. +49 5251 603379} and Wolfgang Karl H\"ardle$^b$}

\vspace{0.2cm}
\centerline{\large $^a$Department of Economics, Paderborn University}
\vspace{0.2cm}
\centerline{\large $^b$School of Business and Economics, Humboldt University Berlin}



\vspace*{0.5cm}

\begin{abstract}

\noindent Penalized spline smoothing  of time series and its asymptotic properties are studied. 
A data-driven algorithm for selecting the smoothing parameter is developed. 
The proposal is applied to define a semiparametric extension of the well-known Spline-GARCH, called a P-Spline-GARCH, based on the log-data transformation of the squared returns.  
It is shown that now the errors process is exponentially strong mixing with finite moments of all orders.  
Asymptotic normality of the P-spline smoother in this context is proved. 
Practical relevance of the proposal is illustrated by data examples and simulation. 
The proposal is further applied to value at risk and expected shortfall.


\vspace{.3cm}


\vspace{.3cm}

\end{abstract}

\newpage

\centerline{\Large\bf A data-driven P-spline smoother}

\vspace{0.3cm}

\centerline{\Large\bf applied to economic and financial time series}

\setcounter{page}{1}

\section{Introduction}


P-spline (penalized spline) regression (Eilers and Marx, 1996) becomes a very popular nonparametric smoothing technique due to its flexibility and computational advantages. The idea of P-splines traces back to O'Sullivan (1986). See also Ruppert et al. (2003). In this paper, P-spline smoothing of time series with a truncated polynomial basis will be investigated. Asymptotic properties of this approach with independent errors are e.g. studied by Wand (1999), Hall and Opsomer (2005), Kauermann (2005), Claeskens et al. (2009) and Xiao (2019). P-spline regression with correlated errors is e.g. discussed in Krivobokova and Kauermann (2007). See also Kauermann et al. (2011). However, asymptotic properties of P-splines under stationary time series errors
and data-driven selection of the smoothing parameter in this context are not yet well studied. 

We first adapted some asymptotics of Claeskens et al. (2009) and Wand (1999), hereafter CKO09 and WMP99, to P-spline regression under stationary errors. It is shown that this approach is consistent under weak regularity conditions and the asymptotic variance is affected by the correlated errors. But, the estimates still achieve the same optimal rate of convergence as in the i.i.d. case. Furthermore, an approximation of the optimal smoothing parameter is obtained following WMP99 and a self-contained IPI (iterative plug-in, Gasser et al., 1991) algorithm for selecting the smoothing parameter is developed. This algorithm is fully nonparametric, where the unknown variance factor is estimated by a data-driven log-window approach (B\"uhlmann, 1996). The proposal is compared to a recently proposed data-driven local cubic regression (Feng et al., 2020). 
%


Both data-driven approaches are applied to estimate a smooth scale function in GARCH (generalized autoregressive conditional heteroskedasticity, Engle, 1982; Bolllerslev, 1986) by means of the log-transformation of the squared returns. This idea provides new, automatically non-negative estimates of the scale function in a Semi-GARCH (semiparametric GARCH) model (Feng, 2004). The case with the P-spline estimator is an extension of the Spline-GARCH of Engle and Rangel (2008) (without exogenous variables) and will be called a P-Spline-GARCH. 
A Semi-GARCH model can be applied to decompose risk measures into a baseline, a conditional and a local component, which helps to improve the estimation quality and forecasting of market risk. For more information on those approaches and their applications, we refer the reader to Amado et al. (2018).

If parametric GARCH models, such as the GARCH, the APARCH (asymmetric power ARCH, Ding et al., 1993) or the EGARCH (exponential GARCH, Nelson, 1991) are fitted to the standardized returns, new variants of the Semi-GARCH, Semi-APARCH and Semi-EGARCH models will be defined. 
It is shown that the errors in the log-data under those models have finite moments of all orders, are strongly mixing with exponential decay and have exponentially decay autocorrelations as well. Moreover, it is shown that the P-spline smoother obtained from the log-data is asymptotically normal.
The practical performance of the proposals is illustrated by different data examples and further confirmed through a simulation study,   
which show that the error in the estimated volatility of a GARCH model can be clearly reduced by the proposed approaches. For instance, for two selected examples, the mean average absolute error of the estimated volatility obtained by a Semi-GARCH model is just about 35\% or 45\% of that of a parametric GARCH model, respectively. 
It is found that the two nonparametric approaches perform quite similarly. The performance of the local cubic regression is slightly better, but the P-spline smoother is more flexible and runs much faster.

Like GARCH models, the Semi-GARCH approaches can be applied to forecast the VaR (value at risk, JPMorgan, 1996) and ES (expected
shortfall, Acerbi and Tasche, 2002), a risk measure with superior aggregation properties to VaR. The former is used in the current Basel Accord and is being replaced by the latter in the forthcoming finalization of  Basel III (see Basel Committee on Banking Supervision, BCBS, 2016, 2017). Application of the Semi-GARCH models in this context is illustrated by two examples of one-day out-of-sample forecasts of VaR and ES calculated at the $99\%$- and $97.5\%$ confidence levels for a period of about one year, as required by the BCBS (2016, 2017). Hence, our proposals provide useful alternatives to parametric GARCH models for forecasting quantitative risk measures. Now, whether Semi-GARCH approaches perform better than parametric GARCH models or not depends on the data under consideration. Detailed discussions on those topics with a comparative study will be carried out elsewhere.

The paper is organized as follows. The P-spline smoother is studied in Section 2. The IPI-algorithm is developed in Section 3. Section 4 studies the application to Semi-GARCH models. The empirical results and simulation study are reported in Sections 5 and 6. Application of Semi-GARCH models to VaR and ES is illustrated in Sections 7. Final remarks in Section 8 conclude. Proofs of the results are put in the appendix.

\section{P-spline smoother for time series}

We now discuss P-spline smoother in nonparametric regression with time series errors. 

\subsection{The additive nonparametric time series model}
Consider the modeling of a time series $y_{t}$, $t=1, ..., n$, with a deterministic trend, the commonly used nonparametric regression model in this context is given by 
\begin{equation}\label{NP-TS}
y_t= m(\tau_t) + \zeta_t,
\end{equation}
where $\tau_t=(t-0.5)/n$ denotes the rescaled time, $m$ is a smooth trend and $\zeta_t$ is a zero mean stationary error process with autocovariances $\gamma_{\zeta}(l)=\cov(\zeta_1, \zeta_{l+1})=E(\zeta_1 \zeta_{l+1})$. 
This paper proposes to estimate $m$ in (\ref{NP-TS}) by a P-spline smoother. Our main focus is on the development of a suitable data-driven algorithm for selecting the smoothing parameter under stationary errors without any parametric assumptions on $\zeta_t$, so that the estimation of $m$ is fully nonparametric. This requires the development of a data-driven nonparametric estimation of the variance factor $c_f$ in the proposed approximation of the optimal smoothing parameter, where $c_f=[\sum_{l=-\infty}^\infty \gamma_{\zeta}(l)]/(2\pi)=f(0)$ and $f(\omega)$ denotes the spectral density of $\zeta_t$. This will be done by adapting the data-driven lag-window estimator of $c_f$ in B\"uhlmann (1996) to the current context under suitable regularity conditions.

\subsection{The proposed P-spline smoother}
The commonly used P-splines are those with the B-Spline basis because of its numerical stability and computational efficiency. In the current paper, we will however use P-splines with a truncated polynomial basis for smoothing time series due to the following reasons: 1) Such a proposal is a direct extension of the quadratic splines used in Engle and Rangel (2008) and 2) The useful approximation of the optimal smoothing parameter in WMP99 is obtained for the truncated polynomial basis. This result will be adapted to P-splines with correlated errors. Previous studies on this topic are e.g. Krivobokova and Kauermann (2007) and Kauermann et al. (2011). Asymptotic results for P-spline regression under independent errors are obtained by Hall and Opsomer (2005), CKO09 and Xiao (2019). 

In the following $K$ equidistant internal knots, $x_1=1/(K+1)$, ..., $x_K=K/(K+1)$ together with $x_0=0$ and $x_{K+1}=1$, will be used. The $p$-th order P-Spline regression estimator is the minimizer of the following penalized least squares  
\begin{equation}\label{PLS}
Q_{PS}=\sum_{t=1}^{n}\left( y_t-\sum_{i=0}^{p}\theta_i \tau_t^i-\sum_{i=1}^{K}\theta_{p+i} (\tau_t-x_i)^p_+\right)^2 + \lambda^{2p}\sum_{i=1}^{K}\theta^2_{p+i},
\end{equation}
where $(\tau_t-x_i)^p_+=\max[0, (\tau_t-x_i)^p]$ are the truncated polynomial splines and $\lambda>0$ is the penalty or smoothing parameter.  Define $Y=(y_1,...,y_n)'$, 
\begin{eqnarray*}
&& T=
\begin{bmatrix}
1 & \tau_1 & \cdots & \tau_1^p & (\tau_1-x_1)_+^p & \cdots &(\tau_1-x_K)_+^p\\
\vdots & \vdots & \ddots & \vdots & \vdots & \ddots & \vdots\\
1 & \tau_n & \cdots & \tau_n^p & (\tau_n-x_1)_+^p & \cdots &(\tau_n-x_K)_+^p\\
\end{bmatrix} \mbox{ and }  \\ & & \\
&& D={\rm diag}\{{\mathbf 0}_{(p+1)\times 1}, {\mathbf 1}_{K\times 1}\} .
\end{eqnarray*} 
The $p$-th order P-spline estimator of $m$ takes the form of a ridge regression given by
\begin{equation}\label{mPS}
\hat m_\lambda= T(T'T + \lambda^{2p} D)^{-1} T'Y ,
\end{equation} 
which is a linear smoother with a symmetric smoother matrix $S_\lambda=T(T'T + \lambda^{2p} D)^{-1} T'$. One clear difference between local polynomial regression and P-spline smoothing is that the former is carried out at all estimation points while the latter is computed over the intervals determined by the knots, which runs usually much quicker. 

In this paper, $\lambda$ defined above following WMP99 will be used during the estimation procedure and in particular in the IPI-algorithm to be developed. However, the main part of the asymptotic discussion is based on CKO09. To simplify the representation of those results, the smoothing parameter defined by those authors, i.e. $\lambda^*=\lambda^{2p}$, will be employed in the asymptotic results adapted from their work. The specifications of $\lambda$ and $\lambda^*$ are equivalent to each other. But the use of $\lambda$ has some numerical advantage. 
Let $q=p+1$. For further discussion we introduce the following regularity conditions. 

{\bf A1.} {\it $m(\cdot)$ is at least $q$-times continuously differentiable, i.e. $m(\cdot)\in C^{q}([0, 1])$.}

{\bf A2.} {\it $\zeta_t$ is a stationary process with absolutely summable acf  $\sum_{l=-\infty}^\infty |\gamma_{\zeta}(l)|<\infty$.}

{\bf A3.} {\it $K$ internal equidistant knots are used, where $K=O(n^{\nu_K})$ with $0\le\nu_K<1.$}

{\bf A4.}  {\it The smoothing parameter is of the order $\lambda^*=O(n^{\nu_\lambda})$ with $0\le\nu_\lambda<1$.}

{\bf A5.}  {\it $K$ and $\lambda^*$ are chosen so that $K_q>1$, where $K_q$ is as defined in (13) of CKO09.}

Condition A1 is a standard assumption in nonparametric regression. 
A2 is a standard assumption in nonparametric regression models with short memory time series errors.
The asymptotic results in CKO09 are mainly obtained based on these results for regression splines in Zhou et al. (1998), under Assumptions 1 to 3 stated therein. A3 in this paper is the same as their Assumption 3, but with equidistant knots. Their Assumptions 1 and 2 are automatically fulfilled by the equidistant design considered here. The condition $\nu_\lambda<1$ in A4 ensures that the shrinkage bias tends to zero, as $n\to\infty$. The other requirement $\nu_\lambda\ge0$ is made for simplicity and is unnecessary. The proposed P-spline smoother can even be estimated consistently for some $\lambda^*\to0$, as $n\to\infty$. However, such a choice is far from the optimal one and is not suggested. Condition A5 ensures that our discussion is within the large number of knots scenario of CKO09. For the equidistant design we have the following approximation $K_q=K_{q, \rm A}[1+o(1)]$ with
\begin{equation}\label{Kqa}
K_{q, {\rm A}} = K\cdot(\lambda^*\pi^{2q})^{1/(2q)}n^{-1/(2q)}.
\end{equation}
The proof of (\ref{Kqa}) is given in the appendix. This result allows us to check whether A5 is fulfilled in practice or not. For fixed $K$ and $\lambda^*$, $K_{q, \rm A}$ is a decreasing function of $n$.

Let $\tau$ denote a generic $\tau_t$. Analogously to  (17) and (18) in CKO09 on the asymptotic bias and the asymptotic variance of P-spline regression with a truncated polynomial basis and a large number of knots such that $K_q\ge1$, we have
\begin{lemma}
Under Assumptions A1 -- A5 the P-spline estimator $\hat m_\lambda$ is consistent with the pointwise asymptotic bias and asymptotic variance of the following orders (of magnitude) 
\begin{equation}\label{Bt}
E[\hat m_\lambda(\tau)] - m(\tau) = O(\delta^{p+1}) + O\{[\lambda^*n^{-1}]^{(p+1)/(2p+1)}\} \mbox{ and }
\end{equation}
\begin{equation}\label{Vt}
\var[\hat m_\lambda(\tau)]  = O\{n^{-1} [\lambda^*n^{-1}]^{-1/(2p+1)}\}\footnote[3]{We are grateful to the authors of this work, who kindly provided us a correction of a typo in their  (18). That is the power $2p/(2p+1)$ in that formula should be $-1/(2p+1)$ as it is given here.}, 
\end{equation}
where $\delta=1/(K+1)=O(K^{-1})$ is the length of the interval between two consecutive knots. 
\end{lemma}
The proof of Lemma 1 is omitted, because those results are the same for models with i.i.d. or short memory stationary errors. Lemma 1 shows that the P-spline smoother is consistent under very weak conditions. The optimal smoothing parameters are $\lambda^*_{\rm opt}=O[n^{2/(2p+3)}]$ and $\lambda_{\rm opt}=O\{n^{1/[(2p+3)p]}\}$, respectively. If $\lambda^*_{\rm opt}=O[n^{2/(2p+3)}]$ and $K=O(n^{\nu_K})$ with $\nu_K\ge1/(2p+3)$ are used, then under A1, A2 and A5, $\hat m_\lambda$ achieves the so-called optimal rate of convergence given by Stone (1982) for any nonparametric regression estimator of a function $m\in C^q[0, 1]$.  If $K_q<1$, the P-spline smoother is still consistent after a slight adjustment of A4. Now, the results reduce to those in (15) and (16) of CKO09 for the scenario with a small $K$. This will not be considered further. 

A crucial problem for the practical implementation of the proposed P-spline smoother is the development of a suitable data-driven algorithm for selecting $\lambda$. Results in Lemma 1 are not suitable for this purpose, because the corresponding constants are unknown. Hall and Opsomer (2005) claimed that the result in their  (35) might be used to develop a plug-in algorithm for selecting $\lambda$. But they also indicated that their formula is too complex to be attractive for practical use. This idea will hence not be considered here. 
In this paper we will follow the idea of WMP99 to obtain a rough approximation for $\lambda_{\rm opt}$ and then use it to develop an IPI-algorithm for selecting $\lambda$. Note that the finite sample MASE (mean averaged squared error) of $\hat m_\lambda$ is given by  
\begin{equation}\label{MASE}
{\rm MASE}(\hat m_\lambda) = \frac{1}{n}\sum_{t=1}^n E\{[\hat m_\lambda(\tau_t)-g(\tau_t)]^2\} = B(\lambda) + V(\lambda), 
\end{equation}
where $B(\lambda)$ stands for the average squared finite sample bias in the MASE and $V(\lambda)$ is the finite sample (global) variance part. According to  (2) of WMP99, we have
\begin{equation}\label{B2}
B(\lambda) = \frac{1}{n}||(S_\lambda-I)m||^2, 
\end{equation}
where $||a||=\sqrt{a'a}$ for a vector $a$. Note that $B(\lambda)$ is not affected by the correlated errors. But, the variance of $\hat m_\lambda$ depends on $c_f$. By means of standard techniques in nonparametric regression with time series errors we obtain the following approximation
\begin{equation}\label{AV}
V(\lambda) \approx \frac{2\pi c_f}{n}tr(S_\lambda^2).
\end{equation}
This adapts the well-known fact in nonparametric regression with correlated errors to the current case. The proof of (\ref{AV}) is given in the appendix. Two changes comparing to the corresponding result under i.i.d. errors are: 1) Instead of the variance of the errors, $V(\lambda)$ is determined by the total sum of all autocovariances of $\zeta_t$ and 2) This result is only an approximation and an exact finite sample formula is now no longer available.

Using the arguments in WMP99 we can obtain the following approximation of $\lambda_{\rm opt}$ 
\begin{equation}\label{LA}
\lambda_{\rm A}=\left( \frac{2\pi c_f tr[(T'T)^{-1}D]}{||T(T'T)^{-1}D(T'T)^{-1}T'm||+2\pi c_f tr\{[(T'T)^{-1}D]^2\}}\right)^{1/(2p)}.   
\end{equation}
A sketched proof of (\ref{LA}) is given in the appendix. See also Kauermann (2005) for similar results obtained under a linear mixed model. 
If the errors are uncorrelated, we have $c_f=\sigma^2_{\zeta}/(2\pi)$ and $\lambda_{\rm A}$ reduces to that given in  (4) of WMP99. On the one hand,  $\lambda_{\rm A}$ is an extension of   (4) in WMP99. On the other hand, our result can be thought of as a simplification of that given in Section 2.6 of WMP99 based on the complete covariance matrix of $\zeta_t$. This simplified result is easy to use in practice. 

A plug-in algorithm for choosing $\lambda$ can be developed based on (\ref{LA}). A great advantage of $\lambda_{\rm A}$ compared to the asymptotically optimal bandwidth in local polynomial regression is that here only a pilot estimate of $m$ itself is required. 
We will see that $\lambda_{\rm A}$ is obtained under the further condition $\lambda^{2p} (T'T)^{-1}=o(1)$, which may not be fulfilled in a practically relevant case. Hence, $\lambda_{\rm A}$ seems to be a biased approximation of $\lambda_{\rm opt}$. 
Nevertheless, WMP99 indicated that this formula is very useful for choosing the smoothing parameter in P-splines under i.i.d. errors. This fact will also be confirmed by the data examples and the simulation in this work. Hence, we propose the use of $\lambda_{\rm A}$.    


\section{The IPI-algorithm for P-splines smoothing}

Our proposal consists of an IPI-algorithm for estimating $c_f$ and a main procedure.

\subsection{The IPI-algorithm for estimating $c_f$}

Denote the estimate in the $i$-th iteration obtained with the smoothing parameter $\lambda_{i-1}$ by $\hat m_{i}$. Let $\zeta_{t, i}=y_t-\hat m_{i}(\tau_t)$ be the corresponding residuals.
The sample autocovariances of $\zeta_{t, i}$, $\hat \gamma_i(l)$ say, can be calculated. Then the variance factor $c_{f, i}$ is estimated through
\begin{equation}\label{cfbw}
\hat c_{f, i}=\frac{1}{2\pi}\sum_{l=-M_i}^{M_i} w_{l, i} \hat \gamma_i(l),
\end{equation}
 where $M_i$ is the window-width (of the lag-window estimator) used in the $i$-th iteration and $w_{l, i}=l/(M_i+0.5)$ are weights calculated using the Bartlett-window. In this paper we propose to select the optimal $M_i$ using the IPI-algorithm of B\"uhlmann (1996) with some minor adjustments, which reads as follows. See also Feng et al. (2020). For this purpose, the following stronger assumptions on the error process instead of A2 are required.  

{\bf A2$'$.} {\it $\zeta_t$ is a stationary process with absolutely summable cumulants up to order 8 and quickly decaying acf such that $\sum_{l=0}^\infty (l+1)^4 |\gamma_{\zeta}(l)|<\infty$. Moreover, let $f^{(1)}(\omega)$ denote the first generalized derivative of $f(\omega)$. It is assumed that $f^{(1)}(0)\ne0$.}

A2$'$ summarizes the regularity conditions for the consistency of the following IPI-algorithm in B\"uhlmann (1996). It implies that $E(\zeta_t^{8+\delta})<\infty$. The relationship between A2$'$ and the commonly used $\alpha$-mixing condition in the current context is also discussed there.  In particular, A2$'$ is fulfilled by an ARMA process with finite eighth moment.     

The proposed IPI-algorithm to estimate $c_f$ reads as follows. 
\begin{enumerate}
\item[\enspace\enspace i)]  Set the initial window-width to be $M_{i, 0}=[n/2]$ with $[\cdot]$ denoting the integer part. 

\item[\enspace ii)]  Global stage: In the $j$-th iteration, $\int[f(\lambda)^2]d\lambda$ is estimated. Then the integral of the first generalized derivative of $f(\lambda)$,  $\int f^{(1)}(\lambda) d\lambda$, is estimated using the Bartlett-window and the window-width $M_{G, j}'=[M_{G, j-1}/n^{2/21}]$. One obtains $M_{G, j}$ by inserting the estimates into  (5) of B\"uhlmann (1996). The procedure will be carried out iteratively until it converges or 20 iterations are achieved. We obtain $\hat M_G$.

\item[iii)] Local adaptation for estimating $c_f=f(0)$:  Calculate $f(\lambda)$ and $\int f^{(1)}(\lambda) d\lambda$ again but using the window-width $M'=[\hat M_G/n^{2/21}]$. Put those estimates into the formula of the local optimal bandwidth at $\lambda=0$ in (5) of B\"uhlmann (1996) to obtain $\hat M_i$.   
\end{enumerate}
The rate of convergence of the resulting $\hat c_f$ is of the order $O_p(n^{-1/3})$, which can be improved further, if a more complex $C^2$-window is used. A minor adjustment we made is that the inflation factor $n^{-2/21}$ instead of $n^{-4/21}$ in the original proposal is used, because, as noted by B\"uhlmann (1996), smaller inflation factors improve the rate of bias. 
A small simulation shows that the adjusted inflation factor works better than the theoretically optimal one. But now, a few more iterations are required. We do not fix the number of iterations but propose to run the procedure until convergence is reached. Furthermore, we propose the use of a larger starting window-width compared to that used in the original procedure so that the resulting estimates are more stable. Indeed, the selected window-width is usually not affected so much by $M_0$, if the procedure converges. Roughly speaking, the procedure can be carried out using any suitable initial window-width. Throughout the above procedure only the Bartlett-window is used. The resulting $\hat \gamma_i(l)$ are $\sqrt{n}$-consistent and the rate of convergence of $\hat c_f$ is not affected by the errors in $\hat m_{i}(\tau_t)$.


\subsection{The main IPI algorithm}

Based on the formula of $\lambda_{\rm A}$, WMP99 proposed a plug-in algorithm for selecting $\lambda$ by means of some pilot procedure and showed that this idea works well in practice. However, we will not use his algorithm directly due to the following reasons: 1. A correlation adaptive pilot procedure is not well studied in the literature. And 2. Even if such a pilot method exists, it is usually based on a search procedure and runs very slowly. If the sample size is very large, like by the data examples considered in this paper, the computing time of an algorithm for selecting $\lambda$ based on a search procedure will be too long. Because of similar reasons, some other ideas for selecting the smoothing parameter in P-splines with a truncated polynomial basis, e.g. those proposed by Kauermann (2005), Krivobokova and Kauermann (2007) and Krivobokova (2013), will also not be considered. 

In this paper we will adapt the well-known IPI-idea of Gasser et al. (1991) for kernel or local polynomial regression to select $\lambda_A$. This IPI-algorithm is similar to Wand's method, but without the use of a pilot rule and runs much faster. The effect of the correlation is considered from the first iteration. The proposed algorithm processes as follows.
\begin{enumerate}
	\item Put the number of knots to be $K=\min(n/4, 40)$ and set $\lambda_0=0.2$.  
		\item In the $i$-th iteration for $i\ge 1$, estimate $\hat m_{i}$ using the smoothing parameter $\lambda_{i-1}$.
	\item Calculate the residuals and estimate $c_f$ using the proposed IPI-algorithm. 
	\item Obtain $\lambda_{i}$ by inserting $\hat m_{i}$ and $\hat c_f$ into (\ref{LA}).
	\item Repetitively carry out Steps 2 to 5 until $|\lambda_i-\lambda_{i-1}|=o(n^{-1})$, for $i>1$, or 20 iterations are achieved. Denote the finally selected smoothing parameter by $\hat\lambda$. 
\end{enumerate}  
Note that both, $K$ and $\lambda_0$, are fixed independently of $n$. We will see that, for $\lambda_0$ from a very large range, $\hat\lambda$ is not affected by the choice of this initial value. Of course the required number of iterations may depend on the distance between $\lambda_{\rm A}$ and $\lambda_0$. The value $\lambda_0=0.2$ is just fixed for simplicity in application. Concerning the choice of $K$, it is found that, if $n$ is large enough, any $K>20$ can be indeed used. In most of the cases, the proposed algorithm is already stable for any $K\ge 10$. However, sometimes the estimation quality could be affected, if $K<20$. Furthermore, it is worth indicating that the proposed IPI-algorithm is stable for any $K$, if it is large enough, and it even works well with $K=n$. The selected $\lambda$ adapts to $K$ automatically, so that the estimation quality is almost not affected. But we do not propose the use of a too large $K$, e.g. with $K>60$, because this will strongly increase the computing time without improving the estimation results. If $K$ is too large, the estimation quality can even become slightly worse. This matter will be further discussed in Section 6.  
The default choice $K=\min(n/4, 40)$ is suggested following Ruppert (2002) and we find that this is a suitable choice for the proposed IPI-algorithm. 
A further requirement on the proposed IPI-algorithm is that A5 must be fulfilled by $K$ and $\lambda_0$ for given $n$. It is easy to show that this condition is fulfilled in any practically relevant case. Some concrete examples will be given in Sections 5 and 6. Moreover, in the implementation in R, $(T'T)^{-1}$ is solved by the generalized inverse, because the common matrix inverse operator usually does not work in this case.

The nice practical performances of this IPI-algorithm will be illustrated in sections 5 to 7 using data examples, through a simulation study and by applying the proposals to VaR and ES. The most important theoretical property of $\hat\lambda$ is stated in the following.
\begin{theorem}
Under the assumptions of Lemma 1, the selected smoothing parameter $\hat\lambda$ by the proposed IPI-algorithm for the P-spline smoother of time series is a consistent estimator of $\lambda_{\rm A}$such that
\begin{equation}\label{T1}
(\hat\lambda-\lambda_{\rm A})/\lambda_{\rm A} = o_p(1). 
\end{equation}
\end{theorem}
The proof of Theorem 1 is omitted. It holds, because both, $\hat c_f$ and $\hat m$ are consistent. The rate of convergence of $\hat\lambda$ to $\lambda_{\rm A}$ will not be discussed in the current paper. Note however that Theorem 1 does not mean that $\hat\lambda$ is a consistent estimator of $\lambda_{\rm opt}$. The proposed IPI-procedure converges very quickly. Only a few iterations are usually required, although the stopping criterion is very strict. Following Lemma 1, $\hat m$ and $\hat c_f$ are already consistent in the first iteration. Several further iterations are needed to reduce the effect of $\lambda_0$.


\section{Application to the Semi-GARCH models}

The proposed data-driven P-spline smoother can be applied to smooth time series in different research areas. In particular, it provides a new tool for smoothing macroeconomic time series. In this paper we will employ this approach to develop new variants of the Semi-GARCH class by means of the log-transformation. 
Let $P_t$ be some stock prices or the values of some financial index, $t=0, 1, ..., n$, and $r_t=\ln(P_t)-\ln(P_{t-1})$ denote the (log-) returns. To analyze the slowly changing local (unconditional) variance and the conditional heteroskedasticity simultaneously, we define the following model
\begin{equation}\label{semig}
r_t =  \mu + \sqrt{v(\tau_t) h_t} \varepsilon_t,
\end{equation}
where $\varepsilon_t$ are i.i.d. random variables with zero mean and unit variance, $v(\cdot)>0$ stands for the local variance function and $h_t$ denotes the conditional variance of $\xi_t=r_t^*/\sqrt{v(\tau_t)}=\sqrt{h_t} \varepsilon_t$ with $r_t^*=r_t-\mu$. 
In this model, $\sqrt{v(\cdot)}>0$ is assumed to be a smooth scale function. The product of the local and conditional variances, $\sigma^2_t=v(\tau_t) h_t$, is called the total variance of $r_t^*$.
It is assumed that $\xi_t$ has a unit variance so that the model is uniquely defined. This implies that $E(h_t)=1$. For the practical implementation we put $\hat\mu=\bar r$. 
 
The above definition provides a wide extension of the Semi-GARCH model (Feng, 2004). If it is further assumed that $\xi_t$ follow a unit GARCH model, we have 
\begin{equation}\label{hgarch}
h_t= \alpha_0 + \sum_{i=1}^p \alpha_i \xi_{t-i}^2 + \sum_{j=1}^q \beta_j h_{t-j},
\end{equation}
where $\alpha_i, \beta_j\ge 0$ with $\sum_{i=1}^q \alpha_i+\sum_{j=1}^p \beta_j<1$. Due to the restriction $E(\xi_t^2)=1$, the scale parameter $\alpha_0=1-\sum_{i=1}^q \alpha_i-\sum_{j=1}^p \beta_j$ is no longer free.  Equations (\ref{semig}) and (\ref{hgarch}) together define a Semi-GARCH with the standard GARCH model in the parametric part. 
If the scale function changes over time, the Semi-GARCH model should be employed. 

Engle (2003) indicated that $h_t$ is a weighted sum of long-, middle- and short-term volatility effects with the weights $\alpha_0$; $\beta_1, ..., \beta_p$; and $\alpha_1, ..., \alpha_q$, respectively. 
A Semi-GARCH model is indeed approximately a GARCH model with a time-varying scale parameter, i.e. with a time-varying weight for the long-term unconditional variance. 
By means of a Semi-GARCH model the total volatility is decomposed into a long-run (local) component $\sqrt{v(\tau_t)}$, a short-run (conditional) heteroskedasticity $h_t^{1/2}$ and the standard deviation of $\varepsilon_t$ as the baseline volatility factor. The estimate $\sqrt{\hat v(\tau_n)}$ reflexes the current volatility level and helps to improve the estimation and forecasting of the volatility.


The function $v(\cdot)$ can e.g. be estimated by kernel or (non-negatively constrained) local linear regression based on the multiplicative model. Following Engle and Rangel (2008), $v(\cdot)$ can also be estimated from the log-transformation of the squared returns. Now, the non-negativity of $\hat v(\cdot)$ is automatically guaranteed. Assume that $\varepsilon_t\ne0$ a.s. (almost sure). 
Further discussions are sometimes given in an a.s. sense without explanation, if no confusion arises. Define $\varsigma_t=\ln(\varepsilon_t^2)$, $\eta_t=\ln(h_t)$,
$y_t=\ln[(r_t^*)_+^2]$, $m(\tau_t)=\ln[v(\tau_t)/C_\varepsilon]$ and $\zeta_t=\ln[C_\varepsilon(\xi_t)_+^2]$, where $\zeta_t$ is the error process with $E(\zeta_t)=0$ and $C_\varepsilon$ is a constant determined by the distribution of $\xi_t$.
Now, $y_t$ shares the additive model (\ref{NP-TS}) and $m(\cdot)$ can be estimated using any well-developed nonparametric regression technique for time series. An estimator of $v(\tau_t)$ is then given by $\hat C_\varepsilon\exp[\hat m(\tau_t)]$. Our proposal in this paper is to estimate $m(\cdot)$ using the P-spline smoother. The approach defined in this way is an extension of the Spline-GARCH and is hence called a P-Spline-GARCH. 

 By means of this approach, an equivalent scale function can be estimated consistently via $m(\cdot)$ under very weak moment conditions and without any parametric assumptions on $\xi_t$.
To obtain a consistent estimator of $C_\varepsilon$ however, the condition $E[\xi_1^4]<\infty$ is still required, which is the same moment condition as required for fitting a parametric GARCH model and is assumed throughout this paper. Having obtained $\hat m(\tau_t)$ and $\hat v(\tau_t)$, the standardized returns $\hat\xi_t$ can be calculated. Any suitable GARCH model, like the standard GARCH, the APARCH or the EGARCH, can be fitted to $\hat\xi_t$.
Now, we will show that $\hat m(\cdot)$ is asymptotically normal under common regularity conditions. For simplicity, the above three models will be considered as explicit examples. But the results of Theorems 2 and 3 below hold in more general cases. The following (sufficient) moment conditions are required, if the original process follows a GARCH or an APARCH model.

{\bf A6.} The pdf of $\varepsilon_1$, $f_{\varepsilon}(x)$ say, is defined according to (1) in Fern$\rm \acute{a}$ndez and Steel (1998), hereafter FS98, but standardized. It is further assumed that $E(\varepsilon_1^4)<\infty$ and $E(\xi_1^4)<\infty$.

Innovation distributions in well-known packages for GARCH models are usually defined following the proposal of Fern$\rm \acute{a}$ndez and Steel (1998), including the $t$- and skewed-$t$-distributions with df$>4$, the normal and skewed normal distributions as well as the $ged$- and skewed $ged$-distributions as special members, where $ged$ stands for the generalized error distribution. They imply that $\varepsilon_1\ne0$ a.s. and $f_{\varepsilon}(0)>0$. Conditions for the existence of higher order moments of the GARCH and APARCH models are studied by He and Ter\"asvirta (1999a, b) and Ling and McAleer (2002), among others. Those conditions are jointly determined by $f_{\varepsilon}(x)$ and the parameters of the model under consideration.  
For the EGARCH model, we need the following adjusted version of A6.

{\bf A6$'$.} The moment generating function (mgf) of $\varepsilon_1$ exists in $(-\delta, \delta)$, for some $\delta>0$, and $\sum\limits_{i=1}^\infty|\beta_i|<\infty$, where $\beta_i$ are the EGARCH coefficients as defined in (2.1) of Nelson (1991). 

The conditions required for the existence of the mgf of $\varsigma_1$ is not changed. The stronger condition in A6$'$ on $\varepsilon_1$ is required so that the mgf's of $\eta_1$ and $\xi_1$ exist, because in an EGARCH, $h_t$ is determined by $e^{\varepsilon_t}$. This condition is fulfilled by the $ged(\nu)$ and the skewed $ged(\nu)$ with a shape parameter $\nu>1$, including the normal and skewed normal distributions with $\nu=2$. Moreover, we assume that $\sum\limits_{i=1}^\infty|\beta_i|<\infty$, instead of  $\sum\limits_{i=1}^\infty\beta_i^2<\infty$ to exclude possible long memory cases. 
For further discussions we still need the strong mixing property of the above models. Related results may e.g. be found in Carrasco and Chen (2002), and Lee and Shin (2004). 
Now, we have  
\begin{theorem}
Let A6 for GARCH and APARCH, and A6$'$ for EGARCH hold, then
\vspace{-0.2cm}
\begin{enumerate}
\item[i)] The mgf's of the log-processes $\varsigma_t$, $\eta_t$ and $\zeta_t$ all exist and hence all of these log-processes have finite moments of all orders. 
\end{enumerate}
\vspace{-0.2cm}
Assume further that the conditions on the mixing properties of GARCH and APARCH, or those on the mixing properties of the EGARCH(1, 1) as given e.g. in Carrasco and Chen (2002), and Lee and Shin (2004), respectively, hold, then
\vspace{-0.2cm}  
\begin{enumerate}
\item[ii)] The log-processes in i) are all $\beta$-mixing with exponentially decay and 
\item[iii)] The acf's of those processes also decay exponentially.   
\end{enumerate}
\vspace{-0.2cm}
\end{theorem}
The proof of Theorem 2 is given in the Appendix. Theorem 2 i) provides some very nice properties of the log-innovation process.
Theorem 2 ii) extends the exponential $\beta$-mixing (and hence geometric ergodic) property of squared GARCH, APARCH and EGARCH to the corresponding log-processes, which reflect the fact that mixing properties of a positive process will not be affected by the log-transformation. For the EGARCH, ii) and iii) are given for a first order model only, because we do not find known mixing properties of general EGARCH($p$, $q$) models. 
The result in Theorem 2 iii) extends a well-known fact for a log-normal process to more general case, which together with those in Theorem 2 i) ensures that assumption A2$'$ required by the data-driven lag-window estimator of $c_f$ is fulfilled. Now, we are ready to prove the asymptotic normality. Let $\hat m_{n, \lambda_n}(\tau)$ denote a sequence of estimates with the corresponding smoothing parameters $\lambda_n$. Furthermore, we introduce the following adjusted assumptions.

{\bf A3$'$.} {\it The number of knots is chosen by $K_n=O(n^{\nu_K})$ with $1/(2p+3)<\nu_K<1/2.$}

{\bf A4$'$.} {\it The smoothing parameter is chosen of the optimal order $\lambda^*_n=O[n^{2/(2p+3)}]$.}

The lower bound of $\nu_K$ in A3$'$ is made so that the approximation bias is asymptotically negligible. The upper bound $\nu_K<1/2$ is taken to simplify the proof of Theorem 3, which ensures $s_{tj}$ in $S_{\lambda_n}$ are all of the order $o(n^{-1/2})$.
A4$'$ means that the chosen penalty term is of the optimal order. According to (\ref{Kqa}), a reasonable choice is $K=n^{1/(2p+2)}$, which satisfies A3$'$ and also implies A5. The constants in the asymptotic bias and variance may depend on $\tau_t$. This will however not affect our proof of the asymptotic normality and will not be considered in detail. Hence, we will denote a generic $\tau_t$ by $\tau$ for simplicity. Under A4$'$, the P-spline smoother achieves the optimal rate of convergence with $B_{n, \lambda_n}=E[\hat m_{n, \lambda_n}(\tau)] - m(\tau)=O(n^{-\nu_m})$ and $\var[\hat m_{n, \lambda_n}(\tau)]\approx n^{-(2\nu_m)} V_{n, \lambda_n}$, where $\nu_m=(p+1)/(2p+3)$ and $V_{n, \lambda_n}$ is a sequence of constants in the finite sample variances. 
A suitable central limit theorem for a weighted sum of a strongly mixing sequence is given by Peligrad and Utev (1997), hereafter PU97, which applies to the proposed P-spline smoother. Following their results we have
\begin{theorem}
Let A1, A3$'$, A4$'$ and A5 hold. Under the conditions of Theorem 2, we have
\begin{equation}\label{AsyN}
n^{\nu_m}[\hat m_{n, \lambda_n}(\tau)-m(\tau)-B_{n, \lambda_n}]/\sqrt{V_{n, \lambda_n}}\stackrel{\cal D}\to N(0, 1).
\end{equation}
\end{theorem}
This result can be extended to the case with $\nu_\lambda < 2/(2p+3)$ together with the assumption that the approximation bias is still asymptotically negligible. Now, asymptotic normal distribution is only determined by the asymptotic variance with a lower rate of convergence. 
Asymptotic normality of a kernel estimator, $\hat v_{\rm K}(\tau)$ say, of the scale function in the Semi-GARCH model obtained from the multiplicative model directly is proved in Feng (2004). Following Theorem 2, his results can be extended to a general class of Semi-GARCH models with an estimate of the scale function obtained independent of the used parametric volatility models in the second stage. However, Theorem 3 is on the asymptotic normality of $\hat m(\tau)$, not on that of $\hat v(\tau)$. It is clear that in the current context the resulting $\hat v(\tau)$ is no longer asymptotically normal but asymptotically log-normal.     

The effect of the errors in $\hat m_\lambda(\tau)$ on the parameter estimation based on the standardized returns can be studied following Theorem 3 in Feng (2004). Two related facts are that the optimal smoothing parameter $\lambda^*_{\rm opt}$ is not the optimal choice for the further parameter estimation and the parameter estimation from the standardized returns with $\lambda^*=O(\lambda^*_{\rm opt})$ is no longer $\sqrt{n}$-consistent. Detailed discussion on this topic is omitted to save space. All results in this section hold for the local polynomial estimator of $m$ obtained from the log-transformation of the squared returns under corresponding conditions.

\section{Real data examples}

In the following, the proposed P-spline smoother is employed to estimate the scale functions in the returns of three indexes, namely the Standard and Poor 500 (S\&P), the DAX 30 (DAX) and the Nikkei 225 (NIK), from January 1988 to April 2018, and of three German companies, i.e. the SAP SE (SAP), the Siemens AG (SIE) and the Allianz SE (ALV), from January 1997 to April 2018, downloaded from Yahoo Finance. 
As indicated by Harvey et al. (1994), the centralized returns are (a.s.) non-zero.  Hence, $\ln[(r_t^*)^2]$ is well defined in practice and there is no computational problem with possible zero returns.
The main results are obtained with the default choices $K=40$ and $\lambda_0=0.2$. 
The obtained results are then compared to those of a data-driven local cubic regression under stationary time series errors (Feng et al., 2020). The latter is well-developed and is used here as a benchmark. The Alg. B there is used as the authors suggested. This comparative study will be further carried out through the simulation study in the next section. As far as we know, this is the first empirical study to compare the performance of data-driven P-splines to that of a data-driven local regression. In the following the abbreviation Semi-GARCH stands for a model with the standard GARCH in the parametric part or for a generic model in this class. That with APARCH in the second part will be called Semi-APARCH. The parametric models will be fitted using the `fGarch' package in R. The Semi-EGARCH will not be considered, because it cannot be fitted by `fGarch'. 

The chosen $\hat\lambda$ for P-splines and the selected bandwidth ($\hat b$) for the local cubic regression for all examples are listed in the first part of Table 1, where the integer in brackets under $\hat\lambda$ indicates the number of required iterations. 
A surprising finding is that, for the chosen $K$, the amount of the smoothing parameter according to the definition in (\ref{PLS}) is at the same level as that of the bandwidth of a local cubic regression. 
Usually, $\hat \lambda$ will increase with $\hat b$. However, the orders of those two quantities are not the same. And the variation of $\hat\lambda$ is clearly larger than that of $\hat b$, indicating that those two smoothing parameters have different features.  
The return series and the estimated scale functions using both approaches are displayed in Figure 1 for the examples of S\&P, DAX and SAP. 
We see, the obtained results using both approaches are quite similar. The differences at the endpoints or around the extrema of a scale function may be sometimes slightly clear. Note that the data-driven local cubic regression achieves the optimal rate of convergence. It hence confirms that the proposed data-driven P-spline smoother works well in practice.           

To show the possible effect of $\lambda_0$ on $\hat\lambda$, the IPI-algorithm with $K=40$ was further carried out with different $\lambda_0$ within a wide range. The results for $\lambda_0=0.05, 0.8$ and 3.2, as well, are given in the second part of Table 1. The biggest differences between $\hat\lambda$ selected using different $\lambda_0$ are all smaller than $n^{-1}$ and are hence negligible. The required number of iterations can be affected by $\lambda_0$ slightly. Moreover, in all cases the condition $K_{q, \rm A}>1$ is clearly fulfilled by $\lambda_0$ and $\hat\lambda_i$ in each iteration. The smallest value of $K_{q, {\rm A}}$ is that for the DAX with $\lambda_0=0.05$. Now, we have $K_{q, {\rm A}}=4.344$, which is clearly bigger than 1. It is also shown that $\hat\lambda$ adapts to $K$ automatically, so that $\hat m$ is almost not affected by the choice of $K$, provided that $K\ge 30$. Those results are omitted to save space. 




\section{A simulation study}
In the second stage, a suitable GARCH model can be fitted to the standardized returns $\hat\xi_t$. Then the total volatility can be calculated.
The aim of the following simulation study is to show whether and how far the estimation quality of the volatility can be improved by a Semi-GARCH approach compared to that of a GARCH model. So far as we know such topics were not yet investigated in the literature. In contrast to most of the simulation studies in this context, the quality of $\hat\lambda$, $\hat m$ or the further resulting parameter estimation from the residuals or standardized returns will not be discussed. Moreover, the two estimated scale functions of S\&P and SAP instead of some artificially designed scale functions will be used as the underlying scale functions. Here, the results of the local cubic regression are used, so that they are independent of the choice of $K$. The two GARCH(1, 1) models under conditional normal distribution with $h_{1t}=0.05 + 0.08 \xi_{t-1}^2+ 0.87 h_{t-1}$ and $h_{2t}=0.10 + 0.13 \xi_{t-1}^2+ 0.77 h_{t-1}$, will be employed as the error processes (in the multiplicative model). Where the coefficients are adjusted slightly so that their sum is exactly 1. The two sample sizes $n_1=7641$ and $n_2=5365$ are used in the two cases, respectively. Therefor, only one sample size in each case is considered.
The simulation was carried out using the local cubic regression (LC) and cubic P-splines (PC) with $K=10$, 20, ..., 70, denoted by PC1, ..., PC7, respectively. In each case $J=1000$ replications were carried out. The parametric GARCH model under constant scale assumption (CS) is taken as a comparison.    

We propose to assess the quality of the estimated volatility by given approach `X' using a goodness-of-fit measure defined in the following. Firstly, we define an error criterion MAAE (mean average absolute error) of the estimated volatility over all replications
\begin{equation}\label{MAAE}
 M_{\rm X} = \frac{1}{J}\sum_{j=1}^J \left[\frac{1}{n}\sum_{t=1}^n |\hat\sigma_{jt, \rm X}-\sigma_{jt}|\right],
\end{equation}
where $\hat\sigma_{jt, \rm X}$ is the estimated (total) volatility series in the $j$-replication by approach `X' and $\sigma_{jt}$ stands for the corresponding true simulated volatility series. 
Denoting by $M_{\rm CS}$ the MAAE obtained by the comparison, a goodness-of-fit measure, called the reduction of the MAAE (RMAAE in $\%$) of approach `X' comparing to $M_{\rm CS}$, is defined by 
\begin{equation}\label{RMAAE}
 R_{\rm X} = (1-M_{\rm X}/M_{\rm CS})*100\%,
\end{equation}
which indicates the amount of the reduced errors in the estimated volatility obtained by a Semi-GARCH approach compared to that of a parametric GARCH model.    

Here, we have obtained $M_{\rm CS}*10000=6.40$ and 19.75 for the two data sets. The means of $M_{\rm X}*10000$, $R_{\rm X}$ (in \%) and the means of $\hat b$ or $\hat\lambda$, respectively, are listed in Table 2. We see, the estimation quality of the volatility for the S\&P index is much higher than that for the SAP stock price. Furthermore, the absolute error in the volatility estimated by a GARCH model can be clearly reduced by a Semi-GARCH approach. If the scale function is estimated by the LC-method, this reduction is roughly 56.3\% for S\&P and 66.0\% for SAP. The improvement in the latter case, when the estimation quality is poorer, is more clear than that in the former case. The results for the PC-approach depend on $K$ slightly. A clear finding is that $K\le 20$ should not be used. For the SAP-example $M$ and $R$ are optimized by $K=30$ and $M$ increases slightly with $K$ for $K>30$. But the results with $K=40$ are almost the same as those with $K=30$. 
For the S\&P-example, $M$ decreases slightly, if $K$ increases. Here, the $K$ value, which minimized $M$, is larger than 70. However, for all $K\ge 30$, we have $R\approx 55.8\%$. That is, the effect of any $K\ge30$ seems to be very limited. The reason for this is that $\hat\lambda$ adapts automatically to the change of $K$ so that the resulting estimates are almost not affected. In both cases the suggested default choice with $K=40$ is suitable. The use of a $K>40$ in the latter case is also unnecessary. In comparison with the LC-approach, the estimation quality of the PC approach with $K=40$ is slightly poorer. But, the computing time of the latter is clearly shorter than one fifth of that required by the LC-approach. Boxplots of the AAE for all replications for the three approaches CS, LC and PC4 are displayed in Figure 2. We see, the reduction in the estimation error made by a Semi-GARCH model is uniformly over all replications. And the maximal estimation error of a Semi-GARCH approach among all replications is still smaller than the minimal error of the GARCH method.  

\section{Forecasting VaR and ES by Semi-GARCH}

Now, we will show how a Semi-GARCH approach can be applied to forecast VaR and ES. For this purpose the models are fitted without the last $K$ returns, which are then used for backtesting the one-day rolling forecasts of VaR and ES. Here, the negative returns, $-r_t$, will be used as the losses, because they correspond to the ratios of the linearized losses. See (2.5) in McNeil et al. (2015). The forecasting of ES-forecasts based on a Semi-GARCH model will be described. The forecasting of VaR is similar. 
For simplicity, it is assumed that the mean of the returns is constant. And only the conditional $t$-distribution will be considered to save space. Furthermore, we simply propose to use the last estimate of the scale function, $\sqrt{\hat v(\tau_n)}$, as the forecast for the unconditional standard deviation for the next year.   
Then the rolling one-day forecasts of ES are given by
\begin{equation}\label{ESt}
{\rm ES}_\alpha(n+k) = -\bar r + \sqrt{\hat v(\tau_n) \hat h_{n+k}} {\rm ES}_{\epsilon, \alpha},  k=1, ..., K, 
\end{equation}
where ${\rm ES}_{\epsilon, \alpha}$ is the ES of a rescaled $t$-distribution with the df=$\nu$ and variance 1. Here, $\hat h_{n+k}$ are the forecasts of the conditional variances following a unit GARCH model, which hence have to be calculated by the descaled returns $r_t^*=(r_t-\bar r)/\sqrt{v(\tau_n)}$, $t=n+1, ..., n+K$. 
According to  (2.25) in McNeil et al. (2015) we have, for $\nu>2$,  
\begin{equation}\label{ESte}
{\rm ES}_{\epsilon, \alpha} = \frac{f_\nu (t_\nu^{-1}[\alpha])}{1-\alpha} \frac{\nu + [t_\nu^{-1}(\alpha)]^2}{\nu -1}\sqrt{\frac{\nu-2}{\nu}}, 
\end{equation}
where $f_\nu$, $t_\nu$ and $t_\nu^{-1}$ are the density, distribution and quantile functions of a $t$-distribution with the df=$\nu$, respectively. Note that ${\rm ES}_{\epsilon, \alpha}$ corresponds to a VaR at the confidence level $\alpha^*$ with $\alpha^*(\nu, \alpha)=t_\nu({\rm ES}_{\epsilon, \alpha}\sqrt{\nu/[\nu-2]})$, which is a function of $\alpha$ and $\nu$. Now, $\alpha^*$ is slightly bigger than 0.99 so that the $97.5\%$-ES is slightly bigger than the $99\%$-VaR.

The above idea is illustrated by calculating 250 out-of-sample forecasts of VaR at both $\alpha$-levels and ES at the $97.5\%$-level for the DAX and SAP, where the model was estimated using P-spline scale function and an in-sample till about April 2017.  
The results with an APARCH(1, 1) in the parametric part are displayed in Figure 3. We see, the estimated $97.5\%$-ES are quite similar to the estimated $99\%$-VaR. The POT (points over the threshold) for DAX and SAP are 3 and 4 for the $99\%$-VaR, and 9 and 11 for the $97.5\%$-VaR, respectively. The first three are within their green zones, i.e. [0, 4] and [0, 10] for $\alpha=99\%$ and $97.5\%$, respectively, and the last lies at the beginning of its yellow zone, according to the Traffic-Light approach proposed by the BCBS for backtesting the VaR. The POT for the $97.5\%$-ES are 2 and 4, respectively, with one point fewer than that for the VaR in the first case. This is as expected, because under the $t$-distribution, the POT for the $97.5\%$-ES is usually equal to, but sometimes smaller than that for the $99\%$-VaR.  
Those examples show that the Semi-GARCH approaches provide useful alternatives to the parametric GARCH models for measuring quantitative risk. It is also found that, although the in-sample performance of a Semi-GARCH approach is usually better than that of a parametric GARCH model, the out-of-sample forecasts of VaR and ES using a Semi-GARCH model are not necessarily better than those obtained by a parametric model. Moreover, the selected bandwidth based on the in-sample is not necessarily the best for calculating the out-of-sample forecasts of VaR or ES. Those problems and related topics, such as the backtesting of VaR and ES, will be studied in more detail elsewhere.

\section{Concluding remarks}

In this paper a P-spline smoother for time series with an IPI-algorithm for selecting the smoothing parameter is proposed without any parametric assumption on the errors. This algorithm is developed based on an approximation of the optimal smoothing parameter obtained following WMP99, which works well in practice. Comparing to the traditional local cubic regression, the P-spline smoother is more flexible and runs much faster.
The proposal is applied to estimate the scale function in a Semi-GARCH model from the log-transformation of the squared returns. This leads to a so-called P-spline-GARCH. Properties of the log-processes and asymptotic normality of the resulting trend estimate are studied in detail. Practical performance of the P-Spline-GARCH is illustrated by a number of examples. A simulation study confirms that the errors in the estimated volatility of a GARCH model can be strongly reduced by means of a Semi-GARCH approach. Moreover, it is shown that the proposals can be employed for forecasting VaR and ES. 
Further studies on the application of Semi-GARCH approaches to VaR and ES are of great interest. It is also worth to extend the current proposal to P-splines with the B-spline basis or to models with long-range dependent errors. Possible semiparametric extensions of long-memory GARCH models, such as the FIGARCH (Baillie et al., 1996) or FIEGARCH (Bollerslev and Mikkelsen, 1996), are also of great interest. 


{\bf Acknowledgments:} The data were downloaded from Yahoo Finance. We are grateful to Prof. Tatyana Krivobokova, University of G\"ottingen, for detailed explanation about their asymptotic results on P-splines with truncated polynomial basis. We would like to thank Prof. Timo Te\"aäsvirta, CREATES, Aarhus University, and Prof. Luo Xiao, North Carolina State University, for sending us their most recent research papers. Our thanks also go to Prof. Thomas Gries, Mr. Marlon Fritz, Mr. Sebastian Letmathe and Mr. Xuehai Zhang, Paderborn University, for helpful discussion and suggestions.  




\newpage

\section*{References}
\vspace{-.5cm}
\begin{description}
\setlength\itemsep{0pt}

\item  Acerbi, C. and Tasche, D. (2002). Expected Shortfall: a natural coherent alternative to Value at Risk. {\it Econ. Notes}, 31, 379-388.


\item Amado, C., Silvennoinen, A. and Ter\"asvirta, T.  (2018). Models with multiplicative decomposition of conditional variances and correlations. Preprint, Aarhus University.

\item Baillie, R., Bollerslev, T. and Mikkelsen, H. (1996). Fractionally integrated generalised autoregressive conditional heteroscedasticity, {\it J. Econometr.}, 74, 3–30.


\item BCBS (2016). STANDARDS: Minimum capital requirements for market risk.   

\item BCBS (2017). Dec 2017 {\it Basel III: Finalising post-crisis reforms}. 



\item Bellingsley, P. (1995). {\it Probability and Measure Theory} (3rd ed.). Wiley, New York.    

\item Bollerslev, T. (1986) Generalized autoregressive conditional heteroskedasticity. {\it J. Econometrics} {31}, 307--327.

\item Bollerslev, T. and Mikkelsen, H.O. (1996).  Modeling  and  Pricing  Long  Memory  in  Stock  Market  Volatility. {\it J. Econometr.}, 73, 151-184.


\item B\"{u}hlmann, P. (1996). Locally adaptive lag-window spectral estimation. \textit{J. Time Ser. Anal.}, 17, 247-270.

\item Claeskens, G., Krivobokova, T., and Opsomer, J. (2009). Asymptotic properties of penalized spline estimators. {\it Biometrika}, 96, 529–544.


\item Carrasco, M., Chen, X., 2002. Mixing and moment properties of various GARCH and stochastic volatility models. Econometric
Theory 18, 17-39.

\item Davis, R. A. and Mikosch, T. (2009).  Probabilistic Properties of Stochastic Volatility Models.
in Andersen, T.G., Davis, R.A., Kreiß, J.-P. and Mikosch, T. (ed): {\it Handbook of Financial Time Series}, pp. 255-267, Springer, Berlin. 

\item Ding, Z., C.W.J. Granger  and  R.F. Engle
(1993) A long memory property of stock market returns and a new
model. {\it J. Empirical Finance} 1, 83-106.

\item Eilers, P.H.C. and Marx, B.D. (1996). Flexible smoothing with B-splines and penalties (with discussion). {\it Statist. Sci.}, 11, 89-121.

\item
Engle, R.F. (1982) Autoregressive conditional heteroskedasticity with estimation of U.K. inflation, {\it Econometrica} {50},
987--1008.


\item Engle, R.F. (2003) Risk and volatility: Econometric models and financial practice. Nobel Lecture,
available at http://nobelprize.org.


\item Engle, R.F. and Rangel, J.G. (2008). The Spline-GARCH model for low-frequency volatility and its global macroeconomic causes. {\it Rev. Financ. Stud.}, 21, 1187-1222. 

\item Feng, Y. (2004). Simultaneously modelling conditional heteroskedasticity and scale change.
{\it Econometric Theory}, 20, 563-596.

\item Feng, Y., Gries, T. and Fritz, M. (2020). Data-driven local polynomial for the trend and its derivatives in time series. Revised for a journal.

\item Fern$\rm \acute{a}$ndez, C. and Steel, M.F. (1998). On bayesian modeling of fat tails and skewness. {\it J. Amer. Statist. Ass.}, 93 359-371.


\item Gasser, T., Kneip, A. and K\"{o}hler, W. (1991). A flexible and fast method for automatic smoothing. \textsl{J. Amer. Statist. Assoc.}, 86, 643-652.



\item Hall, P., and Opsomer, J. (2005). Theory for penalised spline regression. {\it Biometrika}, 92, 105-118.

\item Harvey, A., Ruiz, E. and Shephard, N. (1994). Multivariate stochastic variance models. {\it Rev. Econ. Stud.}, 61, 247-264. 

\item He, C. and Ter\"asvirta, T. (1999a). Forth moment structure of the GARCH$(p, q)$ process. {\it Econometric Theory} {15} 824--846.

\item He, C. and Ter\"asvirta, T. (1999b). Statistical properties of asymmetric power ARCH process. In R.F. Engle and H. White (eds). {\it Cointegration,
Causality, and Forecasting: Festschrift in Honour of Clive W.J. Granger}, 
pp. 462-474. Oxford University Press, Oxford.


\item JPMorgan (1996). {\it RiskMetrics Technical Document} (4th ed). JPMorgan, New York.

\item Kauermann, G. (2005). A note on smoothing parameter selection for penalized spline smoothing. {\it J. Statist. Pl. Infer.}, 127, 53–69.

\item Kauermann, G., Krivobokova, T. and Semmler, W. (2011). Filtering Time Series with Penalized Splines. {\it Stud. Nonl. Dynam. \& Econometr.}, 15, 1-28. 

\item Krivobokova, T. and Kauermann G. (2007). A note on penalized spline smoothing with correlated  errors. {\it J. Amer.  Statist.  Assos.}, 102, 1328-1337.

\item Krivobokova, T. (2013). Smoothing parameter selection in two frameworks for penalized splines. {\it J. R. Statist. Soc. B}, 75, 725-741.

\item Lee, O. and Shin, D. W. (2004). Strict stationarity and mixing properties of asymmetric power GARCH
models allowing a signed volatility. {\it Economics Letters}, 84, 167-173.


\item Li, Y. and Ruppert, D. (2008). On the asymptotics of penalized splines. {\it Biometrika}, 95, 415-436.

\item Ling, S. and McAleer, M. (2002). Necessary and sufficient moment
conditions for the GARCH(r,s) and asymmetric power GARCH(r,s)
models. To appear in {\it Econometric Theory}, 18, 722-729.


\item McNeil, A., Frey, R. and Embrechts, P. (2015). {\it Quantitative Risk Management: Concepts, Techniques and Tools}, 2e. Princeton University Press, 
Princeton.

\item Nelson, D. B. (1991). Conditional heteroskedasticity in asset returns: A new approach. {\it Econometrica}. 59, 347-370.


\item O'Sullivan, F. (1986). A statistical perspective on ill-posed inverse problems. {\it Statist. Sc.}, 1, 502–518. 
 

\item Peligrad, M.  and Utev, S. (1997). Central limit theorem for linear processes. {\it Ann. Probab.}, 25, 443-456.


\item Ruppert, D. (2002). Selecting the number of knots for penalized splines. {\it J. Comput. Graph. Statist.}, 11, 735-757.

\item Ruppert, D., Wand, M.P. and Carroll, R.J. (2003). Semiparametric Regression. Cambridge University Press, Cambridge, UK.


\item Stone, C.J. (1982). Optimal global rates of convergence for nonparametric regression. {\it Ann. Statist.}, 10, 1040-1053.




\item Wand, M.P. (1999). On the optimal amount of smoothing in penalised spline regression. {\it Biometrika}, 86, 936-940.


\item Xiao, L. (2019). Asymptotic theory of penalized splines. {\it Elctr. J. Statist.}, 13, 747-794.

\item Xiao, L., Li, Y., Apanasovich, T.V. and Ruppert, D. (2012). Local Asymptotics of P-splines. Preprint, North Carolina State University.



\item Zhou, S., Shen, X. and Wolfe, D. A. (1998). Local asymptotics for regression splines and confidence regions. {\it Ann. Statist.}, 26, 1760-1782.

\end{description}

\newpage
\setcounter{equation}{0}
\renewcommand{\theequation}{A.\arabic{equation}}

{\bf Appendix. Technical details and proofs of results}

{\bf The proof of (\ref{Kqa}).} For $q=p+1$, $K_q$ in CKO09 reduces to
\begin{equation}\label{Kq} 
K_q=K\cdot(\lambda^*\tilde c)^{1/(2q)}n^{-1/(2q)},
\end{equation}
where $\tilde c=c_1[1+o(1)]$ and $c_1$ is a constant that depends only on $q$ and the design density. In our case the design density is the uniform one on $[0, 1]$ with $f(\tau)\equiv 1$, $\tau\in[0, 1]$. The design points $\tau_t=(t-0.5)/n$ satisfy $\int_0^{\tau_t}f(\tau)d\tau=(2t-1)/(2n)$. Speckman (1985) showed tat now $c_1=\pi^{2q}$. That is $\tilde c\approx \pi^{2q}$.  (\ref{Kqa}) is proved by inserting this into (\ref{Kq}). \hfill $\Diamond$

{\bf The proof of (\ref{AV}).}

Let $S_\lambda=(s_{ti})$ with $t, i=1, ..., n$, then $s_{ti}$ for any $t$ are asymptotically equivalent to some kernel weights. Li and Ruppert (2008) and Xiao et al. (2012) provided the asymptotic equivalent kernels of P-splines with the B-spline basis. The asymptotic equivalent kernels in the current context are obtained by Hall and Opsomer (2005) under a white-noise framework. Following Lemma 1, the asymptotic properties of $\hat m(\tau)$ are asymptotically equivalent to those of a $q$th order kernel regression with automatic boundary correction and the corresponding bandwidth $h_{\tau} = O[(\lambda^* n^{-1})^{1/(2p+1)}]$. 
Hence, all $s_{ti}$ are at most of the order $(n h_\tau)^{-1}$ and vary regularly in the interior part. Note that, $V(\lambda)=\{\sum_{t=1}^n \var[\hat m(\tau_t)]\}/n$. Choose $L_1=O(n h_\tau)$ and $L_2=o(n h_\tau)$, such that $L_2\to\infty$, as $n\to\infty$. Define $A=\{L_1+1, ..., n-L_1\}$ and $B_i=\{j: |i-j|<L_1\}$ for given $i\in A$. For given $t$, $i\in A$ and $j\in B_i$, we have $s_{tj}=s_{ti}[1+o(1)]$ and the following decomposition 
\begin{eqnarray*}
\var[\hat m(\tau_t)] & = & \sum_{i=1}^n s_{ti} \sum_{j=1}^n s_{tj} \gamma_\zeta(i-j)\\
& = & \sum_{i\in\bar A}^n s_{ti} \sum_{j=1}^n s_{tj} \gamma_\zeta(i-j) + \sum_{i\in A} s_{ti} \sum_{j=1}^n s_{tj} \gamma_\zeta(i-j) \\
& =: & T_1 + T_2.
\end{eqnarray*} 
For given $i\in A$, the second sum in $T_2$ splits up into two terms
\begin{eqnarray*}
 \sum_{j=1}^n s_{tj} \gamma_\zeta(i-j) & = & \sum_{j\in B_i} s_{tj} \gamma_\zeta(i-j) +  \sum_{j\in\bar B_i} s_{tj} \gamma_\zeta(i-j)\\
& =: & T_3 + T_4.
\end{eqnarray*} 
Straightforward analysis results in $T_3\approx s_{ti} 2\pi c_f$ and $T_4=o(T_3)$. This leads to 
\begin{equation}
T_2 \approx 2\pi c_f\left(\sum_{i\in A} s_{ti}^2\right). \nonumber 
\end{equation}
Furthermore, it can be shown that 
\begin{equation}
T_1 \approx 2\pi c_f O\left[\left(\sum_{i\in\bar A} s_{ti}^2\right)\right]  \nonumber
\end{equation}
and $T_1=o(T_2)$. We obtain
\begin{equation}\label{vartf}
\var[\hat m(\tau_t)] \approx 2\pi c_f\left(\sum_{i=1}^n s_{ti}^2\right) .
\end{equation} 
Result in  (\ref{AV}) is proved by inserting (\ref{vartf}) into $V(\lambda)$. \hfill $\Diamond$

{\bf A sketched proof of (\ref{LA}).}

The proof of this result is similar to that given in WMP99 by replacing $\sigma^2$ there with $2\pi c_f$. Hence, we will only provide some complements about the two approximations used in his proof.
Firstly, the derivation of $\lambda_{\rm A}$ begins with the Taylor expansion of 
$$\left[I + \lambda^{2p} D (T'T)^{-1}\right],$$ where $I$ is the identity matrix. 
The result in (\ref{LA}) is then obtained by assuming that the approximation bias is asymptotically negligible. The latter approximation is ensured by the assumptions of Lemma 1.  
However, the assumption $\lambda\to 0$ he used for the Taylor expansion is unnecessary and also unreasonable, because in the current case $\lambda$ should tend to $\infty$. What we need here is indeed $\lambda^{2p} D (T'T)^{-1}=o(I)$. 
Note that the order of $ T'T$ is $O(n)$, only if $K$ is fixed. When $K\to\infty$, the order of entries of this matrix changes from the first column to the last, and from the first row to the last, as well. The smallest elements occur in the low-right corner. It is easy to show that the last diagonal element is of the order $O(n K^{-(2p+1)})$ and the other elements in that corner are also of this order. Hence, the largest terms of $(T'T)^{-1}$ occur in the low-right corner. Detailed analysis shows that those elements are of the order $O(n^{-1} K^{2p+1})$. Thus, the required condition for the Taylor expansion is indeed $\lambda^{2p} n^{-1} K^{2p+1}=o(1)$ or $\lambda^* n^{-1} K^{2p+1}=o(1)$. Further proof of this result follows the arguments in WMP99. \hfill $\Diamond$

{\bf Remark A.1.} {\it The necessary condition $\lambda^* n^{-1} K^{2p+1}=o(1)$ we obtained is very strong. In particular, it is not fulfilled, if $\lambda^*$ is of the optimal order $O[n^{2/(2p+3)}]$ as obtained by CKO09 with $K=O(n^\nu_K)$ and $\nu_K\ge 1/(2p+3)$, because now $\lambda_{\rm opt}^* n^{-1} K^{2p+1}$ is at least of the order $O(1)$. The condition $\lambda^{2p} D (T'T)^{-1}=o(I)$ is usually also not fulfilled by a practically relevant data set. Hence, $\lambda_{\rm A}$ seems to be only a suboptimal approximation of $\lambda_{\rm opt}$. The search for a better approximation of $\lambda_{\rm opt}$ is an important open question.}

{\bf Proof of Theorem 2.} i) We first prove the results for the innovation process $\varsigma_t=\ln(\varepsilon_t^2)$, which is the same in all of the three models. 
Denote the pdf's of $\varepsilon_1$, $\varepsilon_1^2$ and $\varsigma_1$ by $f_{\varepsilon}(x)$, $f_{\varepsilon^2}(z)$ and $f_{\varsigma}(y)$, respectively.
The proof will be first given an original distribution proposed by FS98. 
Following Equation (1) in FS98, $f_{\varepsilon}(x)$ is defined based on a pdf $f_{s}(s)$ in $(-\infty, \infty)$, which is unimodal and symmetric around 0, such that $f_s(s)=f_s(|s|)$ and the latter is decreasing in $|s|$. A scalar parameter $\gamma\in(0, \infty)$ is added to this distribution to introduce asymmetry. Let $f_s^+(s)=f_s(s)\bbo_{[0, \infty)}(s)$ and $f_s^-(s)=f_s(s)\bbo_{(-\infty, 0)}(s)$. Then $f_{\varepsilon}(x)$ is defined by
\begin{equation}\label{feps}
f_{\varepsilon}(x) = \frac{2}{\gamma+\gamma^{-1}}\left[f_s^+(\gamma^{-1} x) + f_s^-(\gamma x)\right], 
\end{equation} 
where $x\ge0$, if and only if $s\ge0$. The pdf of the two-to-one transformation $\varepsilon_1^2$ is given by 
\begin{equation}\label{feps2}
f_{\varepsilon^2}(z)= \frac{1}{(\gamma+\gamma^{-1})\sqrt{z}}\left[\gamma^{-1}f_s^+(\gamma^{-1} \sqrt{z}) + \gamma f_s^-(-\gamma \sqrt{z})\right]
\end{equation} 
for $z>0$. Now, insert the transformation $y=\ln(z)$ further into $f_{\varepsilon^2}(z)$, after some simplification we have
\begin{equation}\label{fy}
f_{\varsigma}(y)= \frac{e^{y/2}}{\gamma+\gamma^{-1}}\left[\gamma^{-2}f_s^+(\gamma^{-1} e^{y/2}) + \gamma^2f_s^-(-\gamma e^{y/2})\right]
\end{equation} 
A6 implies that $f_{s}(s)=o(|s|^{-5})$ as $|s|\to\infty$, which results in further $f_{\varsigma}(y)=O(e^{y/2})$ as $y\to-\infty$ and $f_{\varsigma}(y)=o(e^{-2y})$ as $y\to\infty$.
Thus, $M_{\varsigma}(u)=E(e^{u\varsigma})$ at least exists for $u\in(-\delta_1, \delta_2)$ with $0<\delta_1<0.5$ and $0<\delta_2\le2$ and $\varsigma_1$ has finite moments of all orders, where $M_{\varsigma}(u)=E[(\varepsilon^2)^u]$ almost surely. 

Moreover, it is easy to show that the above results are not affected by a linear transformation $\tilde\varepsilon_t=a+b\varepsilon_t$ with $b\ne0$, including the standardization as a special case. The reasons for this are, in the above proof: 1) The unimodal property is not required; 2) If there is a mode at zero is unnecessary; and 3) Whether $f_{\varepsilon}(x)$ is increasing for $x<0$ or not, or decreasing for $x>0$ or not, is indeed also not required in the proof given above. Thus those results are preserved after a necessary standardization.

Now, consider $\eta_t=\ln(h_t)$ in GARCH and APARCH. Assume that the initial values all started from their invariant measures. Following the definition of a GARCH or an APARCH model, we have $h_t>\alpha_0$ and the pdf of $h_1$, $f_h(u)$ say, is zero for $u\le\alpha_0$. Hence, $E(h_t^{-k})<\infty$ for any $k>0$. This together with the above proof shows that the mgf of $\eta_t$ at least exists in $(-\infty, 2]$.  This shows that that the mgf of $\zeta_1$ exits, too. The two processes $\eta_t$ and $\zeta_t$ also have finite moments of all orders.   
The proof of the results for $\eta_t=\ln(h_t)$ in the EGARCH under A6$'$ is straightforward and is omitted, because $\eta_t$ is now a linear process. 


ii) Furthermore, because the strong mixing property is defined via $\sigma$-fields and the log-transformation is measurable and almost surely well defined. 
Hence, the mixing properties and the rate of the mixing coefficients of the original process will be all taken over by the log-transformed process. See Davis and Mikosch (2009). The strong mixing properties of the squared GARCH, APARCH and EGARCH models with  exponentially decaying mixing coefficients as given e.g. in Carrasco and Chen (2002), and Lee and Shin (2004) hold immediately for the corresponding log-processes. That is, all of those log-processes are strongly mixing with exponential decay.

iii) The relationship between the mixing coefficients $\theta_k$ of a strongly mixing stationary process $X_t$ with $E(|X_t|^{2+\delta})<\infty$, $\delta>0$, and its autocorrelations $\rho_X(k)$ is given by $$|\rho_X(k)|\le c\theta_k^{\delta/(2+\delta)}, \,\, k\ge0,$$
where $c>0$ is some constant. See Davis and Mikosch (2009) and references therein.  
Hence, $\rho_X(k)$ also decay exponentially, if $\theta_k$ decay exponentially. In the current case, we have further $|\rho_X(k)|\le c\theta_k, \, k\ge0,$ because the processes under consideration all have finite moments of all orders. These results are very helpful for analyzing non-negative processes after the log-transformation. This finishes the proof of Theorem 2. \hfill $\Diamond$

{\bf Proof of Theorem 3.} Most of the results on the asymptotic normality of partial sums or of the sample mean under different strong mixing and moment conditions, such as that in Bellingsley (1995),  are not directly applicable to the proposed P-spline smoother. However, one of the central limit theorems given in PU97 can be applied to a general linear estimator, including the P-spline smoother or the local polynomial regression estimator considered in this paper as special cases. 

Denote by $S_{\lambda_n}$ the sequence of the smoother matrices, we have $$\hat m_{n, \lambda_n}(\tau)=\sum\limits_{i=1}^n s_{n, i, \lambda_n} y_i,$$ where $s_{n, i, \lambda_n}$ are the weights of the P-spline smoother in the corresponding row of $S_{\lambda_n}$. 
Let $\sigma_{n, \lambda_n}^2$ be the finite sample variance of $\hat m_{n, \lambda_n}(\tau)$. Define $\tilde s_{n, i, \lambda_n}=s_{n, i, \lambda_n}/\sigma_{n, \lambda_n}$ and $S_n=\sum_{i=1}^n \tilde s_{n, i, \lambda_n} \zeta_i$. Then we have $E(S_n)=0$ and $\var(S_n)=1$. Now, we only need to show that $S_n$ is asymptotically standard normally distributed, because, 
$$\hat m_{n, \lambda_n}(\tau)-m(\tau)-B_{n, \lambda_n}=\sum\limits_{i=1}^n s_{n, i, \lambda_n} \xi_i$$ by definition and hence, 
$$n^{\nu_m}[\hat m_{n, \lambda_n}(\tau)-m(\tau)-B_{n, \lambda_n}]/\sqrt{V_{n, \lambda_n}}=S_n [1+o_p(1)].$$
The remaining proof is hence to show that Condition (2.1) on the weights $\tilde s_{n, i, \lambda_n}$, Condition (2.2) on the uniform integrability of $\zeta_t^2$ and Condition (c) required by Theorem 2.2 in PU97 on the mixing property and finite moments of $|\zeta_t|$ are all fulfilled by $S_n$. 

Note first that $\zeta_t$ is a strictly and weakly stationary process with finite moments of all orders, and strongly mixing with exponentially decay mixing coefficients. Also $\var(\zeta_t)>0$ is non-degenerate. Hence, Condition (c) of Theorem 2.2 in PU97 is clearly fulfilled. Because of the strong and weak stationarity, $\{\zeta_t^2\}$ forms a uniformly integrable family. Condition (2.2) in PU97 is satisfied. Moreover, the proposed P-spline smoother is asymptotically equivalent to some kernel estimator of order $q=p+1$. But the explicit forms of such asymptotic kernels are very complex and are seemingly not yet well studied in the literature for P-spline smoothers with a truncated polynomial basis. To overcome this difficulty, we hence introduced a slightly stronger restriction on the order of magnitude of the number of knots in A3$'$, which simplifies our proof very much, but is of course unnecessary. Note that the order of $\max\limits_{1\le i \le n}(s_{n, i, \lambda_n})$ is bounded by the magnitude order of the weights of regression splines with $\lambda_n\equiv0$. Under A3$'$ we have $$\max\limits_{1\le i \le n}(s_{n, i, \lambda_n})=O(K_n/n)=O[n^{-(1-\nu_K)}]=o(n^{-1/2}).$$ Under A4$'$ we have $\sigma_{n, \lambda_n}\approx n^{-(p+1)/(2p+3)}\sqrt{V_{n, \lambda_n}}$. Thus, $\tilde s_{n, i, \lambda_n}=o(1)$. This implies $\max\limits_{1\le i\le n}|\tilde s_{n, i, \lambda_n}|\to0$. The standardization condition $\var(\sum_{i=1}^n \tilde s_{n, i, \lambda_n} \zeta_i)=1$ together with the formula of the asymptotic variance ensures that $\sum_{i=1}^n \tilde s_{n, i, \lambda_n}^2=(2\pi c_f)^{-1}[1+o(1)]$, i.e. $\sum_{i=1}^n \tilde s_{n, i, \lambda_n}^2=O(1)$. This means that $\sup\limits_{n}\sum_{i=1}^n \tilde s_{n, i, \lambda_n}^2<\infty$. Condition (2.1) in PU97 is fulfilled by the standardized weights $\tilde s_{n, i, \lambda_n}$. Thus   
$$\sum_{i=1}^n \tilde s_{n, i, \lambda_n} \zeta_i\stackrel{\cal D}\to N(0, 1).$$
Theorem 3 is proved. 
\hfill$\Diamond$

\clearpage
\newpage

\begin{table}
\begin{center}
\caption{Selected smoothing parameters and bandwidths in all cases}
\vspace{0.2cm}
  \begin{tabular}{c|c|ccc|ccc}
  \hline
$\lambda_0$ & $K$ & S\&P & DAX & NIK & SAP & SIE  & ALV \\
\hline
\multicolumn{2}{c|}{$n$} & 7641 & 7660 & 7457 & 5365 & 5444 & 5433  \\
\hline
0.20 & 40 & 0.1607  & 0.2289 & 0.2314 & 0.1753 & 0.1551 & 0.2140\\ 
 & & (5) & (6) & (5) & (5) & (4) & (4) \\ 
 \hline
\multicolumn{2}{c|}{loc. cub. ($\hat b$)} & 0.1811  & 0.2188 & 0.2222 & 0.1966 & 0.1840 & 0.2046  \\  
 \hline \hline
0.05 & \multirow {6}{*}{40} & 0.1607 & 0.2289 & 0.2314 & 0.1753 & 0.1551 & 0.2140 \\ 
  & & (5) & (7) & (7) & (5) & (4) & (5) \\
 0.80 & & 0.1607 & 0.2290 & 0.2314 & 0.1753 & 0.1551 & 0.2141 \\
& &  (7) & (7) & (7) & (7) & (6) & (6)  \\
3.20 &  & 0.1607 & 0.2290 & 0.2314 & 0.1753 &  0.1551 & 0.2141 \\
& & (7) & (7) & (7) & (7) & (6) & (6) \\
\hline
  \end{tabular}
\end{center}
\end{table}

\begin{table}
\begin{center}
\caption{The means of $M_{\rm X}*10000$, $R_{\rm X}$ (in \%) and the means of $\hat b$ resp. $\hat\lambda$}
\vspace{0.2cm}
  \begin{tabular}{c|c|c|ccccccc}
  \hline
Data & Stat & LC & PC1 & PC2 & PC3 & PC4 & PC5 & PC6 & PC7 \\
\hline
\multirow{3}{*}{S\&P} &  $M$ &  2.7924 & 3.0019 & 2.8454 & 2.8225 & 2.8207 & 2.8195 & 2.8187 & 2.8181 \\
& $R$ & 56.258 & 52.932 & 55.404 & 55.760 & 55.787 & 55.805 & 55.817 & 55.826 \\
& $\bar{\hat b}$ & 0.1550 & 0.1272 & 0.1413 & 0.1415 & 0.1474 & 0.1523 & 0.1565 & 0.1601 \\
\hline
\multirow{3}{*}{SAP} &  $M$ & 6.7182 & 6.9481 & 6.9593 & 6.9158 & 6.9161 & 6.9165 & 6.9169 & 6.9173 \\
& $R$ &  65.963 & 64.794 & 64.736 & 64.956 & 64.954 & 64.952 & 64.950 & 64.948 \\
& $\bar{\hat b}$ & 0.1973 & 0.1685 & 0.1887 & 0.1844 & 0.1924 & 0.1990 & 0.2046 & 0.2095 \\
\hline
  \end{tabular}
\end{center}
\end{table}

\clearpage
\newpage

\begin{figure}[!htb]
 \centering
  \includegraphics[bb=0.7cm 0.2cm 21cm 19.5cm, width=20cm, height=20cm, clip]{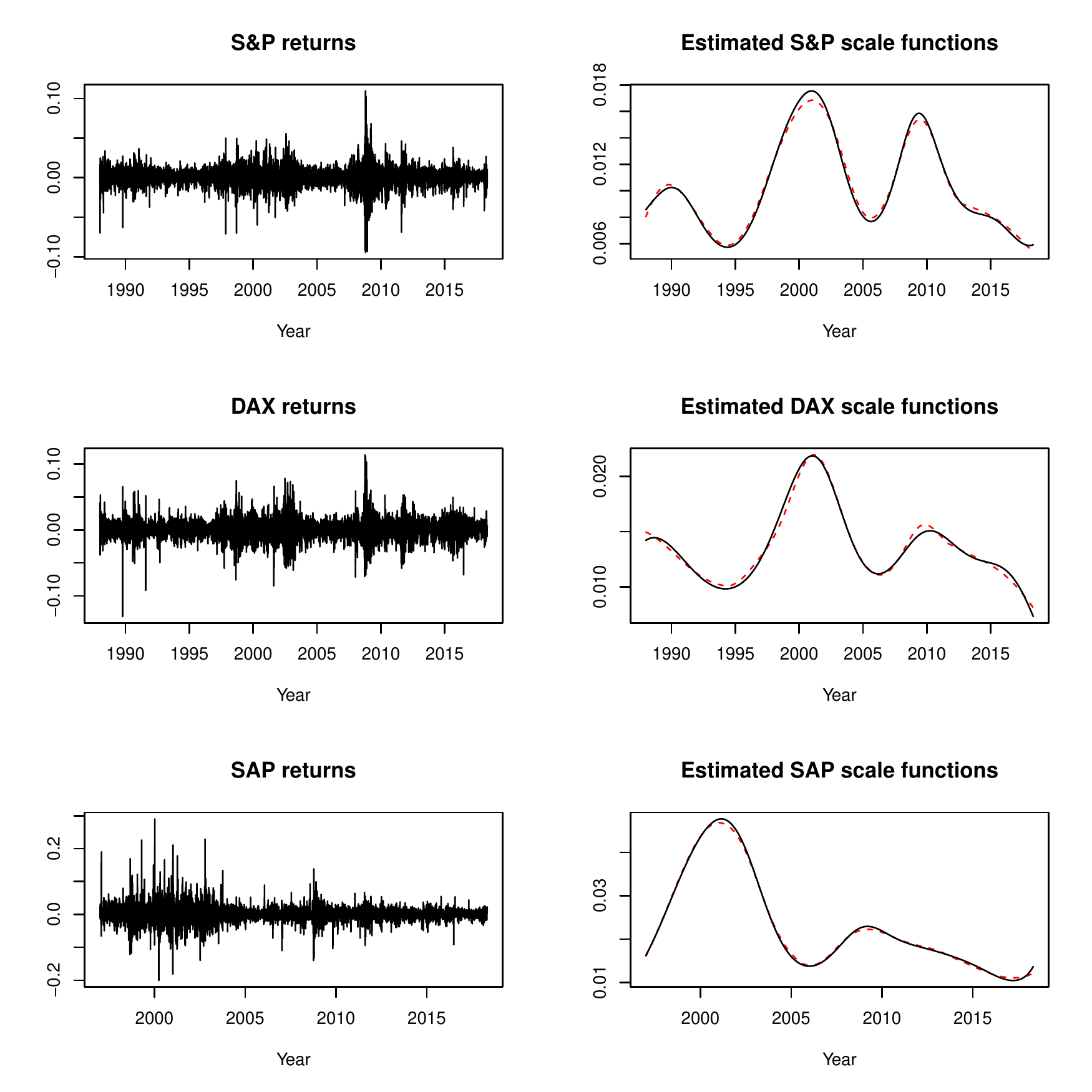}
\caption{Returns (left) for S\&P, DAX and SAP with scale functions (right) estimated by P-splines (solid line) and local cubic regression (dashes).}
  \end{figure}



\begin{figure}[!htb]
 \centering
  \includegraphics[bb=0cm 10.25cm 21cm 19.5cm, width=20cm, height=9cm, clip]{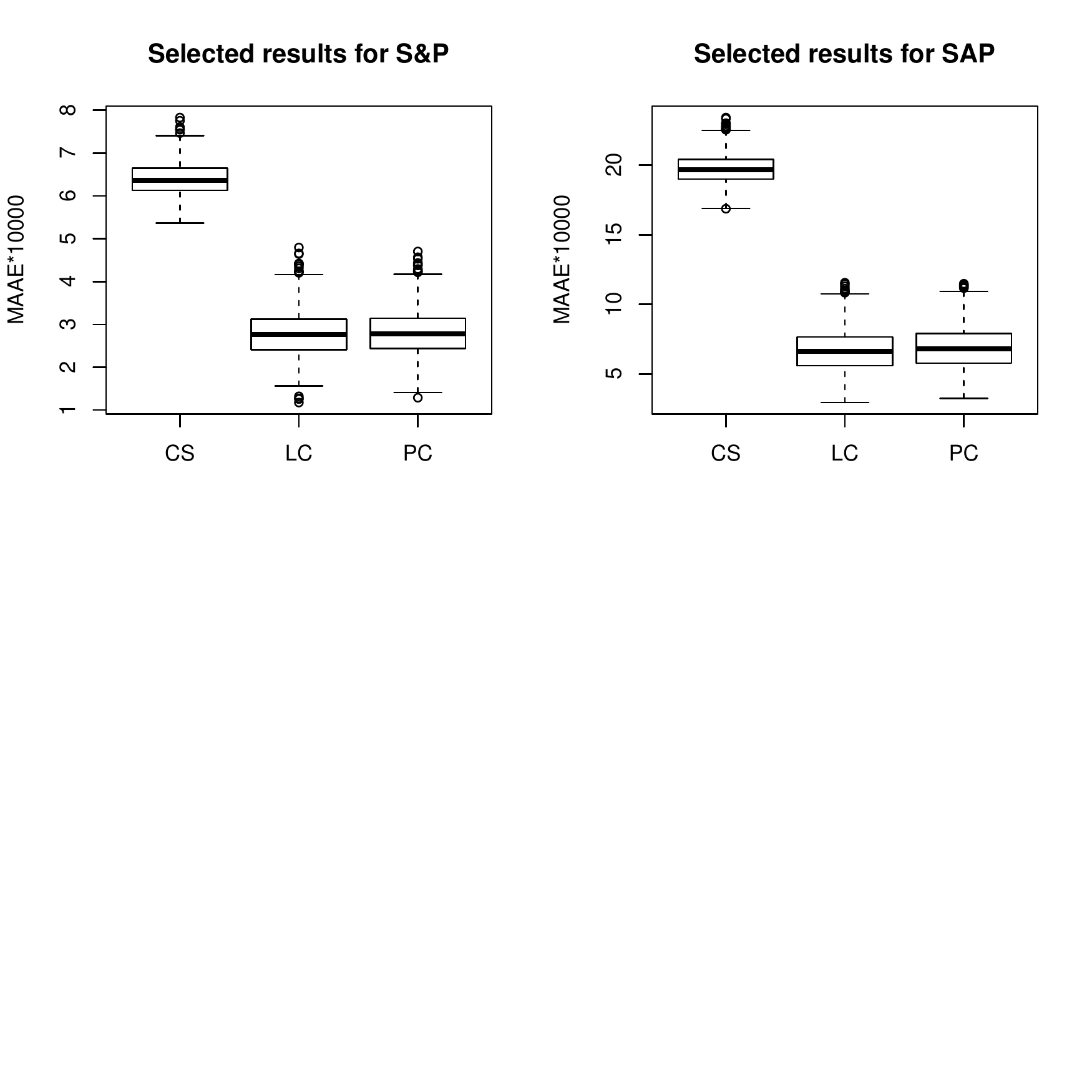}
\caption{Box-plots of MAAE*10000 for CS, LC and PC4 in all replications.}
  \end{figure}

\clearpage
\newpage

\begin{figure}[!htb]
 \centering
  \includegraphics[bb=0.7cm 6.3cm 21cm 19.5cm, width=20cm, height=14cm, clip]{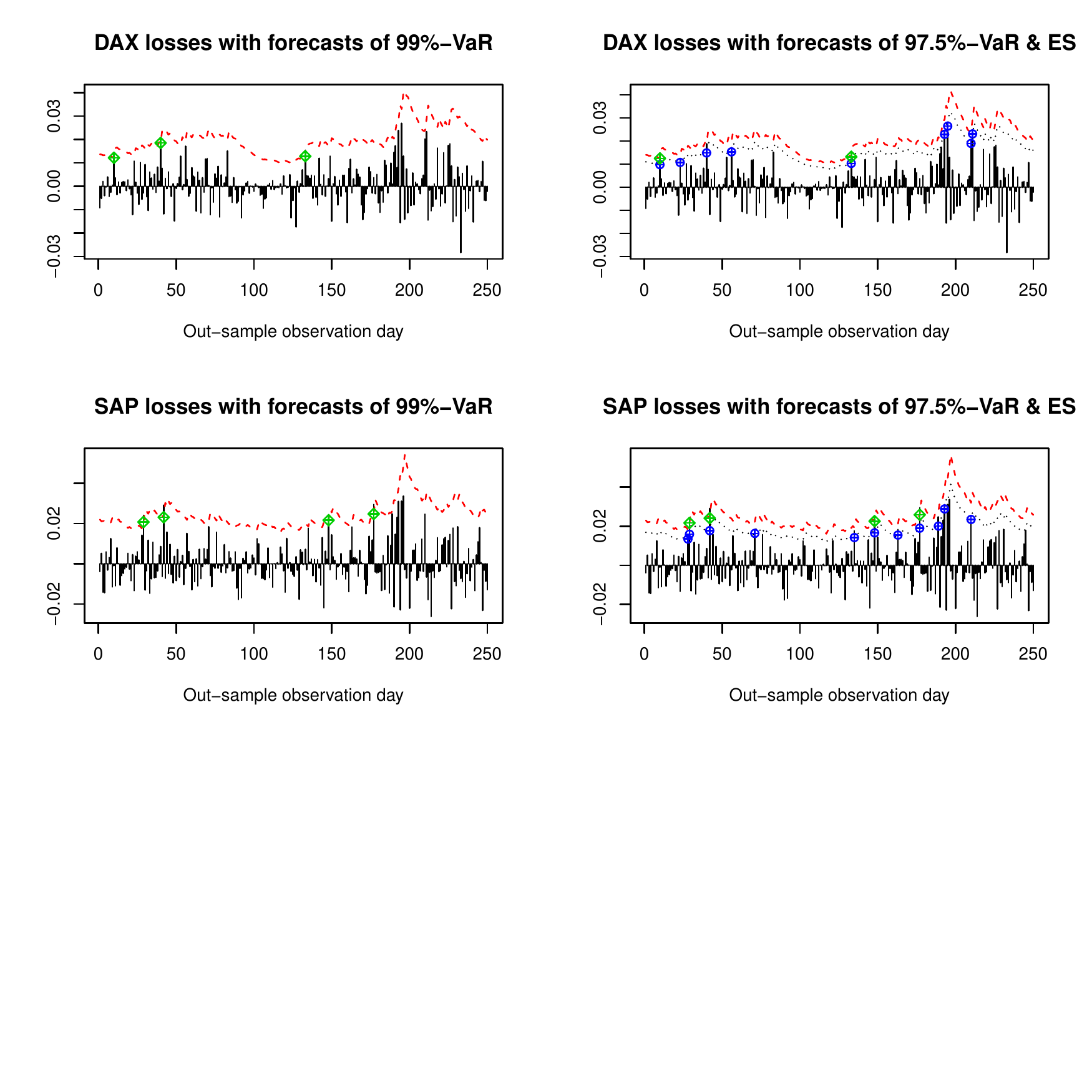}
\caption{Negative returns together with one-day forecasts of the $99\%$-VaR (left), and the $97.5\%$-VaR and ES (right) for DAX (above) and SAP (below).}
  \end{figure}

\end{document}